\documentclass[aps,prl,reprint,superscriptaddress]{revtex4-1}
\usepackage{blindtext}
\usepackage[english]{babel}
\usepackage[utf8]{inputenc}
\usepackage{amsmath}
\usepackage{amsthm}
\usepackage{amsfonts}
\usepackage{graphicx}
\usepackage[colorinlistoftodos]{todonotes}
\usepackage{geometry}
\usepackage{pgfplots}
\usepackage{subcaption}
\usepackage{float}
\usepackage[ruled,vlined]{algorithm2e}
\usepackage{verbatim}
\begin{document}
\title{Port-Hamiltonian inspired gradient diffusion model for magnetohydrodynamic turbulence}
\author{Benjamin Beck}
\email{beck@tu-berlin.de}
\affiliation{Zentrum f\"ur Astronomie und Astrophysik, ER 3-2, Technische Universit\"at Berlin, Hardenbergstr. 36a, 10623 Berlin, Germany}
\author{Wolf-Christian M\"uller}
\email{wolf-christian.mueller@tu-berlin.de}
\affiliation{Zentrum f\"ur Astronomie und Astrophysik, ER 3-2, Technische Universit\"at Berlin, Hardenbergstr. 36a, 10623 Berlin, Germany}

\begin{abstract}
As a reduced representation of the nonlinear spectral fluxes of ideal invariants in incompressible magnetohydrodynamics, we
construct a gradient-diffusion network model that combines
phenomenological considerations and geometrical analysis of the exact
nonlinear energy transfer function. The reduced-order representation
of the conservative spectral transport of energy and cross-helicity
is of port-Hamiltonian form, which highlights the
flexibility and modularity of this approach. Numerical experiments with
Reynolds numbers up to $~10^6$ yield clear power-law signatures of inertial-range energy spectra.
Depending on the dominant timescale of energy transfer, Kolmogorov (-5/3), weak-turbulence (-2), or
Iroshnikov-Kraichnan-like (-3/2) scaling exponents are observed. Anisotropic turbulence in a mean magnetic field is successfully modelled as well.
The characteristic exponents of turbulence decay and the observed influence of cross-helicity on the energy transfer
are consistent with the literature and in agreement with theoretical results.
\end{abstract}

\maketitle

\section{Introduction}
In classical physics, an important open problem is the comprehensive
theoretical understanding of turbulent fluid flows. The highly
nonlinear character of the partial differential equations governing
the fluid velocity field renders the solution deterministically
chaotic.  The statistical properties of the resulting quasi-stochastic
fluid motion, however, displays specific characteristics of order
such as approximate self-similarity, spatio-temporal multi-fractality,
spontaneous stochasticity and dynamic self-organization, see
e.g. \cite{frisch:book,eyink_etal:fluxfreezebreakdown}. As a
consequence, turbulent flows can exhibit super-diffusive transport
\cite{salazar_collins:dispersionrev}, the emergence of structural
coherence \cite{samelson:LCSrev}, and efficient dissipation
\cite{bruno_carbone:solwindlab}, which are processes of fundamental
importance for a large variety of natural phenomena and technical
applications ranging from the formation of stars
\cite{scalo_elmegreen:ismturbrev1} over the dispersion of nutrients in
coastal waters \cite{peters_marrase:planktonturb} to the performance
of turbo-machinery \cite{ashpis_gatski_hirsh:book}.  Fully-developed
realistic turbulence in nature typically exhibits fluctuations of
velocity and other advected quantities with very broad spectral
bandwidth spatially parameterized by the dimensionless Reynolds number
$\mathsf{Re}=L_0V_0/c_d\sim (L_0/\eta)^{4/3}$, see e.g.
\cite{ishihara_gotoh_kaneda:4096sim}.  Here, $L_0$ and $V_0$ are
characteristic length and velocity scales and $c_d$ stands for a scalar
small-scale dissipation coefficient such as kinematic viscosity, $\mu$, or
electrical resistivity, $\lambda$, which quantify small-scale dissipation and yield the
classical Reynolds number, $\mathsf{Re}$, and for plasma systems carrying electrical currents
the magnetic Reynold number, $\mathsf{Rm}$, respectively. The quantity
$\eta=(\mu^3/\varepsilon)^{1/4}$ is the Kolmogorov dissipation length
scale \cite{pope:book} that estimates where nonlinear turbulence
dynamics starts to be dominated by diffusive small-scale dissipation with
$\varepsilon$ denoting the energy dissipation rate.

The strong nonlinearity of turbulence has severely been limiting
progress in the analytical theory of its intricate and complex
statistical properties. Thus, direct numerical simulations (DNS),
which evolve a numerical approximation of the turbulent fields in
discretized space and time, have been a promising way forward during
the last five decades.  Reaching realistic \textsf{Re}, thus, boils
down to obtaining a high enough spatial and temporal resolution of the
numerial model.

However, the Reynold number of atmospheric and oceanic flows can reach
$10^8$ while astrophysical settings allow for even higher values of
\textsf{Re}.  Therefore, in light of the obvious limitations of
computational resources, the DNS approach is not expected to yield
turbulence representations that are even moderately realistic. Heroic
efforts of very high resolution DNS \cite{federrath_etal:10000},
albeit not solving the \textsf{Re}-problem face another challenge:
complexity. The numerical model of reality becomes unwieldy, due to
the sheer amount of generated data.  Thus, increased realism might
thwart the original goal of obtaining better insight into the problem.

In this paper, we present an approach of model reduction that regards
the first-principles of nature as authoritative constraints to our
model of turbulence without the aim of reproducing most aspects of
realistic turbulence with optimal quality.  Instead and closer to the ansatz followed
by scalar shell models
\cite{biferale:shellrev,plunian_stepanov_frick:shellrev,ditlevsen:book}
of turbulence, we focus on specific properties of turbulence dynamics
deemed by us to be fundamentally important while neglecting other
aspects.  The aim of this approach is to isolate specific processes in
an accessible and manageable reduced network model of spectral
transport that allows for straightforward and intuitive manipulation
while respecting fundamental properties determined by the underlying
differential equations of fluid dynamics.

Counter-intuitively, a good starting point for our spectral network
model of turbulence is not two-dimensional incompressible hydrodynamics
but rather three-dimensional incompressible magnetohydrodynamics
(MHD), a fluid approximation of electrically conductive media such as
plasma, e.g. ionized gases, for deeply subsonic flow velocities that
in addition to the velocity, $\mathbf{v}$, also features the magnetic
field, $\mathbf{b}$ \cite{biskamp:book}. In spite of this complication, the MHD approximation as opposed to the
hydrodynamic Navier-Stokes model exhibits an additional symmetry that
straightforwardly justifies the network approach, as shown below in
more detail.  Our approach is aimed at modelling turbulent spectral
fluxes of ideal invariants, i.e. the nonlinear exchange in Fourier
space of quantities such as total energy per unit mass,
$E=1/(2V)\int_V\text{d}V(v^2+b^2)$ and cross-helicity,
$H^\text{C}=1/(2V)\int_V\text{d}V\mathbf{v}\cdot\mathbf{b}$, of a
finite-size, triply periodic system of volume V \cite{biskamp:book3}.
For conceptual and technical simplicity, the present work neglects a
third invariant, magnetic helicity,
$H^\text{M}=1/(2V)\int_V\text{d}V\mathbf{a}\cdot\mathbf{b}$ with the
magnetic vector potential $\mathbf{a}$,
$\mathbf{b}=\nabla\times\mathbf{a}$, which is an entirely magnetic
quantity quantifying topological properties of $\mathbf{b}$
\cite{moffat:maghel}.

In fully-developed turbulence, idealized by the limit
$\textsf{Re}\rightarrow\infty$, neither small-scale dissipation, nor
large-scale effects such as boundary conditions, finite size, or the
energy supply required for the stationarity of this disspative system,
are of importance in the \emph{inertial range} of scales $2\pi/L_0\ll
\ell\ll 2\pi/\eta$, cf., e.g., \cite{frisch:book}. For the respective
Fourier-space wave numbers $k\sim\ell^{-1}$ nonlinear interactions
between Fourier modes give rise to directed mean spectral transport,
e.g. in the case of three-dimensional MHD the direct cascades of total
energy or cross helicity from large to small scales.  This transport
is the statistical consequence of a superposition of many unit
interactions which exchange the quantities in either direction in
wave number following complex principles engrained in the respective
complex nonlinear transfer functions that follow directly from the
primitive MHD eqations \cite{rose_sulem:turbrev}. An essential
hypothesis of the present work is that, on the level of the mean
fluxes of ideal invariants, the intricate detailed interaction
dynamics are not decisive. Instead, we will adopt a much simpler
gradient diffusion ansatz (see below).

The primary goal of our model is a correct representation of the
spectral mean fluxes and the related statistical observables,
e.g. spectral distribution, of total energy and cross helicity. The
motivation for this approach is the construction of an economic
surrogate model of MHD turbulence that can be used in physically more
comprehensive numerical models of such phenomena. In those
simulations, often the complex details of turbulent flows, such as
intermittency and multi-fractality, are not deemed to be of principal
importance as compared to the transport characteristics of the ideal
invariants of the system at hand.

The present paper is structured as follows: In Section II the basic
fluid dynamical equations and their relevant characteristics with
regard to turbulent spectral transport are introduced, followed by the
construction and discussion of the reduced network model and its
natural links to the class of port-Hamiltonian transfer models. In
Section IV basic tests are presented that show the ability of the
model to successfully reproduce various different characteristics of
MHD turbulence and, in particular, its robustness, cost-efficiency,
correctness within the anticipated frame of application.

\section{Fundamental Equations}
In Els\"asser formulation with $\mathbf{z}^\pm=\mathbf{v}\pm\mathbf{b}$
\cite{elsaesser:mhd}, the equations of incompressible,
three-dimensional ideal MHD in Fourier space, read:
\begin{equation}
\partial_t \mathbf{z}^{\pm}(\mathbf{k}) = \mp i \mathbf{k}\cdot \mathbf{b_0} \mathbf{z}^\pm (\mathbf{k}) - i \sum_{\mathbf{k}+\mathbf{p}+\mathbf{q}=0} \mathbf{k} \cdot \mathbf{z}^\mp (\mathbf{p}) \mathcal{P}_{\mathbf{k_\perp}} (\mathbf{z}^\pm (\mathbf{q}))\,. \label{els1} 
\end{equation}
Here, the convolution sum on the right-hand-side (RHS) represents the
basic nonlinearity of the system.  Generally, as is well-known (see,
e.g., \cite{pope:book}), this quadratic nonlinearity results in finite
coupling between groups of three involved Fourier modes whose wave
vectors form a triangle, $\mathbf{k}+\mathbf{p}+\mathbf{q}=0$.  The
operator
$\mathcal{P}_{\mathbf{k_\perp}}(\mathbf{x})=\mathbf{x}-\frac{\mathbf{x}\cdot
\mathbf{k}}{\vert \mathbf{k} \vert^2}\mathbf{k}$ ensures
incompressibility of the evolving Els\"asser fields,
i.e. $\mathbf{k}\cdot\mathbf{z}^\pm=0$, via a projection of the fields
to a plane perpendicular to $\mathbf{k}$.  As the pressure is
generally a nonlinear function of the velocity field, the assumption
of incompressbility, although only justified for subsonic plasma
flows, yields a strong simplification of the problem, turning $p$ into
a passive quantity.  From Eq. \eqref{els1} we obtain the relation
governing the Els\"asser energies, $E^\pm =|\mathbf{z}^\pm|^2/4$ of
the mode with wave vector $\mathbf{k}$:
\begin{equation}
\partial_t {E}^{\pm}(\mathbf{k}) =-\mathcal{I}\bigg( \sum_{\mathbf{k}+\mathbf{p}+\mathbf{q}=0} (\mathbf{k} \cdot \mathbf{z}^\mp (\mathbf{p})) (\mathbf{z}^\pm (\mathbf{q})\cdot \mathbf{z}^\pm (\mathbf{k}))\bigg)\,. \label{els2} 
\end{equation}
From \eqref{els2} it is straightforward to prove that the $E^\pm$ are
conserved individually, such that the nonlinearity in \eqref{els2} can
only result in a transfer of $E^\pm$ between like-signed Els\"asser
fields. This transfer is algebraically determined by the RHS of
Eq. \eqref{els2} - the conservative transfer function.  The constraint
imposed by the conserved Els\"asser energies does not exist on the same
level for Navier-Stokes hydrodynamics and justifies the representation
of nonlinear turbulent transfer as the result of many pair
exchanges. The third mode, present in the divergence term in
\eqref{els2} does not participate in the transfer. Instead, as
suggested by its physical dimension ([time]$^{-1}$), it determines the
timescale of the transfer.

On a geometric level, nonlinear energy transfer in a triad is, thus,
characterized by two structurally discernible contributions to the
transfer function: one determined by the relative
orientation of the corresponding wave vectors and Fourier mode
components, the other setting the timescale of interaction via the
mediator mode and its orientation with regard to the wave-vector triad.
The incompressibility constraint restricts the three-dimensional
Els\"asser modes to the planes orthogonal to their respective
wave vectors.  Therefore, the timescale of nonlinear interaction is
strongly influenced by the shape and respective roles of the edges of
the wave-vector triangle, as well.  In addition, a mutual exchange of the
interacting partners within a given triad changes the sign and not the
magnitude of the energy transfer, consistent with energy
conservation. The interpretation of each Fourier mode as a node in a network,
where mutual links emerge dynamically by conservative Els\"asser energy
transfer, yields an abstract representation of the convolution sum in
Eq. \eqref{els2} and, thereby, facilitates model order
reduction as described below.

However, the complex detailed temporal evolution within one triad, the
collective dynamics of larger triad ensembles, and the relation of
these truncated finite-size structures in Fourier space with regard to
turbulent fluid dynamics in configuration space is hard to establish
\cite{Moffatt2013}. Other triad focused abstractions of nonlinear
turbulence dynamics based on the mechanical analogon of rigid body
rotation \cite{Waleffe1991} and its relation to Jacobi-elliptic
functions \cite{BerionniGurcan2011} have not yet resulted in
substantial complexity reduction.

Triad interactions have also been studied and analyzed regarding to
their stability \cite{Waleffe1992,Linkmann2016}.  These investigations
show an average loss of energy for particular \lq unstable\rq\ triad
modes.  Pseudo-invariants conserved in certain triad geometries
\cite{Rathmann2017} have been identified as well. Although those works
can ascribe different energy cascade mechanisms to specific triad
geometries, it appears to be very challenging to exploit such detailed
information for an improved understanding of physical phenomena
involving turbulence and to
incorporate these insights in reduced-order models of turbulent transport, mixing or
dissipation.

As mentioned in the introduction, the main hypothesis underlying the present work states that the
involved mathematical and dynamical details of Fourier triads are not
decisive for an approximation that delivers a physically consistent
reduced representation of the spectral energy dynamics of
turbulence. Such a reduced-order model of turbulent energy transfer
should be an attractive surrogate model for the application in complex
numerical models in which turbulence is of importance as a physical
ingredient, e.g., regulating energy transport and dissipation.

In the present work, we consider the average flux of energy within the
model network which is determined by the state of its nodes and their
associated wave vectors. The nodes act as reservoirs of quantities
which are conservatively exchanged over the network. This transfer
model is based on additional assumptions which, for example, express
the dynamical tendency towards an asymptotic spectral energy
distribution and the trend to evolve towards minimal-energy states of maximum cross-helicity ($\mathbf{v}\parallel\mathbf{b}$)
or maximum absolute magnetic helicity ($\nabla\times\mathbf{b}\parallel\mathbf{b}$) while accounting for first-principles expressed by the
structure of the MHD equations.

For clarity, let us restate the two key assumptions imposed on the reduced-model ansatz:
i) The focus of our model is on the mean spectral energy dynamics - the detailed complex, non-linear interactions of Fourier modes
expressed as fluctuations on short timescales are assumed to be of minor importance for the statistical energy characteristics.
Instead, the trend towards asymptotic  minimal-energy states is supposed to govern the overall evolution.
ii) Complementary to the fluctuation amplitudes, the geometric structure of the individual nonlinear triad interactions determines
the effectiveness of transport jointly with the dominating nonlinear timescale, neglecting quasi-stochastic phase information.

\section{Port-Hamiltonian network}
The detailed conservation of energy of the Els\"asser-Fourier system
is expected for $\textsf{Re},\textsf{Rm}\rightarrow \infty$ and
motivates a Hamiltonian formulation of the dynamics in the
reduced-order model.  Dissipative small-scale effects, such as
viscosity or resistivity and external influences sustaining the
dynamics of the dissipative system via continuous large-scale energy
input are negligible on the intermediate scales of the inertial range,
but they establish spectral boundary conditions that drive the mean flow of energy from large to small scales as
realized by the above-mentioned conservative transfer dynamics.
Therefore, the inclusion of those non-ideal effects is essential for a
consistent description of turbulence and also for the application of
the proposed model.  For this purpose, we apply a port-Hamiltonian
(pH) \cite{vanderSchaft} framework offering these extensions in a stringent
mathematical form and with the potential of further model reduction. A
matrix formulation, sufficient to describe the Els\"asser MHD system
is an input-state-output pH system of the following form
\cite[p. 101]{vanderSchaft}:

\begin{align}
\partial_t \mathbf{z} &=[J(\mathbf{z})-R(\mathbf{z})]\cdot\frac{\partial H}{\partial
  \mathbf{z}}(\mathbf{z})+\mathbf{u}\,,\\
	{\mathbf{z}}_{out} &=\frac{\partial H}{\partial \mathbf{z}}(\mathbf{z})\,,
\end{align}
with the complex Els\"asser state vector $\mathbf{z}=(\mathbf{z}_{Re}^+
(\mathbf{k}),\mathbf{z}_{Im}^+ (\mathbf{k}),\mathbf{z}_{Re}^+
(\mathbf{q}),\mathbf{z}_{Im}^+ (\mathbf{q}),\mathbf{z}_{Re}^-
(\mathbf{p}),\mathbf{z}_{Im}^- (\mathbf{p})) \in \mathbb{R}^{6\cdot
3}$ defining a minimal self-consistent, conservative set of variables, with the index labels $Re$ and $Im$ indicating real and complex parts, respectively.
The adjoint $J^*$ of the transfer operator $J$ representing the transfer function for
this node fulfills $J^*=-J$ and
the linear dissipation operator $R$ is self-adjoint with non-negative
spectrum. For the standard scalar product, these are given by 
matrix functions with $J(\mathbf{z})=-J(\mathbf{z})^T$ and a positive,
semi-definite $R(\mathbf{z})=R^T(\mathbf{z})\geq 0$.
The large-scale driving of the turbulence is represented by the external port
$\mathbf{u}$.
The operator $H$
is a Hamiltonian function describing the ideal invariants, which are,
for the present MHD system, the Els\"asser energies $E^\pm$ defined
with the complex Els\"asser fields. The Hamiltonian fulfills the passivity property \cite{Mehrmann2019},
i.e. it conservatively re-distributes energy among the six available energy reservoirs, and thus
\begin{align}
H(t_2,\mathbf{z}(t_2))-H(t_1,\mathbf{z}(t_1))\leq \int_{t_1}^{t_2} (
\mathbf{u}(t),{\mathbf{z}}_{out}(t))\text{dt } \end{align} with input
$\mathbf{u}$ and the collocated output ${\mathbf{z}}_{out}$ with the corresponding energy defined by the scalar product $(\mathbf{u},\mathbf{z}_{out})$, and the points in time $t_2\geq t_1$.
From a modelling point of view, we benefit from this pH-formulation in
two ways: First and due to the structure preserving nature of this ansatz, a connection
of pH-systems through their input and output variables forms a larger
pH-system. Second, the formulation is closely related to the
abstract network interpretation of detailed conservative energy
transfer of the original physical system. 

As a result, a single triad interaction, the smallest ideal
pH-subsystem \footnote{A further reduction to three real and complex
components to reduce the system's complexity is
possible, but not employed in this work.}, consists of one summand of
the Els\"asser convolution sum with the Hamiltonian
\begin{align*}
	H(\mathbf{z})=&\frac{1}{2}(\vert \mathbf{z}_{Re}^+ (\mathbf{k}) \vert^2 +\vert \mathbf{z}_{Im}^+ (\mathbf{k}) \vert^2+\vert \mathbf{z}_{Re}^+ (\mathbf{q}) \vert^2\\ &+\vert \mathbf{z}_{Im}^+ (\mathbf{q}) \vert^2+\vert \mathbf{z}_{Re}^- (\mathbf{p}) \vert^2 +\vert \mathbf{z}_{Im}^- (\mathbf{p}) \vert^2)\,,
\end{align*}
and the transfer operator
\begin{widetext}
	\begin{align}
	J(\mathbf{z})=
	\begin{bmatrix}
	0 & 0&  -\frac{1}{2} (\mathbf{k}-\mathbf{q}) \cdot \mathbf{z}_{Im}^- (\mathbf{p}) &-\frac{1}{2} (\mathbf{k}-\mathbf{q}) \cdot \mathbf{z}_{Re}^- (\mathbf{p})&0&  0 \\
	0 & 0&  -\frac{1}{2} (\mathbf{k}-\mathbf{q}) \cdot \mathbf{z}_{Re}^- (\mathbf{p}) &\frac{1}{2} (\mathbf{k}-\mathbf{q}) \cdot \mathbf{z}_{Im}^- (\mathbf{p})&0&  0 \\
	-\frac{1}{2} (\mathbf{q}-\mathbf{k}) \cdot \mathbf{z}_{Im}^- (\mathbf{p}) & -\frac{1}{2} (\mathbf{q}-\mathbf{k}) \cdot \mathbf{z}_{Re}^- (\mathbf{p}) & 0&0&0&0 \\
	-\frac{1}{2} (\mathbf{q}-\mathbf{k}) \cdot \mathbf{z}_{Re}^- (\mathbf{p}) & \frac{1}{2} (\mathbf{q}-\mathbf{k}) \cdot \mathbf{z}_{Im}^- (\mathbf{p}) & 0&0&0&0 \\
	0&0&0&0&0&0\\
	0&0&0&0&0&0
	\end{bmatrix}, 
	A=&
	\begin{bmatrix}
	\mathbf{k} & 0 & 0 &0 &0 &0 \\
	0& \mathbf{k} & 0 & 0 &0 &0 \\
	0 & 0& \mathbf{q}  &0 &0 &0 \\
	0 & 0&0 & \mathbf{q}  &0 &0 \\
	0 & 0&0 &0 & \mathbf{p}  &0 \\
	0 & 0&0 &0 &0 &\mathbf{p} 
	\end{bmatrix}. \label{J}
	\end{align}
\end{widetext}
The skew-symmetry of $J$ becomes obvious in this formulation by
replacing $\mathbf{p}$ with $(\mathbf{k}-\mathbf{q})$, by using
the solenoidality condition $\mathbf{k}\cdot\mathbf{z}^\pm(\mathbf{k})=0$ and the triad relation
$\mathbf{k}+\mathbf{q}+\mathbf{p}=0$. Thus, a triad (port-)Hamiltonian
system
\begin{align}
\partial_t \mathbf{z} &= J(\mathbf{z})\cdot\frac{\partial H}{\partial \mathbf{z}}\cdot(\mathbf{z})+A^T \mathbf{\lambda}\,,\\
A\cdot\mathbf{z}&=0\,,
\end{align}
is introduced, where $\frac{\partial H}{\partial
\mathbf{z}}(\mathbf{z})=\mathbf{z}$ and $\lambda$ is a Lagrange
multiplier ensuring the incompressibility constraint enforced by the
projection operator in \eqref{els1}. The multiplier can be eliminated by a
restriction on the non-compressible submanifold, e.g. by helical (or
Beltrami) decomposition \cite{Constantin1988} of the Els\"asser fields. The entries of $J$
are the individual summands of the triad interaction and define a network of interchanging agents when the single triad formulation of \eqref{J} is extended by a chosen set of Fourier modes and their corresponding dynamics. Additionally, the triad relation
could be dropped from an energetic point of view due to the
skew-symmetric structure, however, conservation of magnetic helicity,
although not considered here, would then be violated.

Dissipative effects are added to the system by $R$. For magnetic
Prandtl number $Pr_m=\frac{Rm}{Re}=1$, dissipation is straightforwardly deduced to be a
diagonal matrix with entries $R_{kk}(\mathbf{z})=\frac{\hat{\mu} +
\hat{\nu}}{2}\vert \mathbf{k} \vert^2$. In the case of unequal
viscosity and resistivity constants, however, it is necessary that
both Els\"asser signs for each wavevector are represented in the
system, since their dissipation effects influence each other due to
their kinetic and magnetic nature. Defining a miniature system of only
one wave number $\mathbf{k}$ with nodes $\mathbf{z}^+(\mathbf{k})$ and
$\mathbf{z}^-(\mathbf{k})$ (real or imaginary part) the dissipation
matrix is given by \begin{align} R=\vert \mathbf{k} \vert^2
\begin{bmatrix} \frac{\hat{\mu} + \hat{\nu}}{2}& \frac{\hat{\mu} -
\hat{\nu}}{2}\\ \frac{\hat{\mu} - \hat{\nu}}{2}& \frac{\hat{\mu} +
\hat{\nu}}{2} \end{bmatrix} \label{eq:diss}\end{align} with eigenvalues
$\lambda_1=\vert \mathbf{k} \vert^2 \hat{\mu} $ and $\lambda_2=\vert
\mathbf{k} \vert^2 \hat{\nu}$, which are both non-negative. Consequently, this holds for the entire system with both
Els\"asser variables present, and hence, ensures positive definiteness
while the symmetry $R=R^T$ is obviously fulfilled.

Since the influence of the mean magnetic field does not change the
Els\"asser energy reservoirs, its dynamics can be added to the skew
symmetric transfer matrix function $J$. Its action on mode
$\mathbf{z}^\pm(\mathbf{k})$ is given by $\mp
i\mathbf{k}\cdot\mathbf{b_0}\mathbf{z}^\pm(\mathbf{k})$, see
\eqref{els1}. This cross-connects real and complex field
components of opposite Elaesser signs consistent with the individual
conservation of the two Els\"asser energies.  Physically, magnetic
perturbation about the mean magnetic field generate oscillations which
continuously exchange energy between the kinetic and magnetic
reservoirs of the Els\"asser variables corresponding to the
propagation of shear Alfv\'en waves along the magnetic field lines
(see e.g., \cite{biskamp:book}). The matrix formulation of the
evolution of
$\mathbf{z}=(\mathbf{z}^+_{Re}(\mathbf{k}),\mathbf{z}^+_{Im}(\mathbf{k}))$
due to $\mathbf{b_0}$ is, then, given by the matrix $B$
\begin{align}
\partial_t \mathbf{z}=B\cdot\mathbf{z}=\begin{bmatrix}
0&\mathbf{k}\cdot\mathbf{b_0}\\
-\mathbf{k}\cdot\mathbf{b_0}&0
\end{bmatrix}\mathbf{z}.
\end{align}
Conservation of the ideal invariants is ensured by the skew-symmetry
of $B$ and, when added to $J$, fills the zero entries one-off the
diagonal that represent the interaction of real and complex parts
within a Fourier mode or a network node. The sum of two skew-symmetric
matrices is still skew-symmetric and therefore, we have formulated a
dissipative network pH system with anisotropy
induced by a mean magnetic field.

More complex networks are obtained by addition of further wave vectors
and corresponding interactions up to the full dynamics represented by
all summands of the convolution sum. Concerning the numerical
implementation, however, the number of interactions scales
quadratically per spatial dimension with the number of nodes.
Consequently, the numerical simulation of a spectrally fully-resolved
system that is not severely truncated becomes unfeasible for even
moderate Reynolds numbers. Thus, such models require an extreme sparse
resolution of wave number space similar to the geometric scaling of
shell models \cite{plunian_stepanov_frick:shellrev}, which partition
wave number space in exponentially growing segments covering many
orders of magnitude with a linearly increasing set of variables. We
exploit the abstraction level of our model by the use of spherical and
cylindrical coordinate systems with exponentially increasing radii
easily obtaining Reynolds numbers above $10^6$ -- the chosen modelling
ansatz allows to combine the benefits of the geometrically discretized
world of low-dimensional shell models with a full three-dimensional
representation that allows anisotropic conservative dynamics.

\section{Gradient Diffusion Network}
A triad interaction is characterized by an inherent periodicity of its
evolution encoded in the dot-product of complex amplitude-vectors and
the corresponding wave vectors present in the transfer functions. As
mentioned above, the Els\"asser energies exhibit detailed conservation
and the activity of a triad is strongly influenced by its geometrical
characteristics. In turbulence, a vast number of triads interact
nonlinearly, resulting, on average, in a direct cascade of energy from
the large, energy-containing scales, towards the small scales, where
dissipative effects become dominant, transforming kinetic and magnetic
energy into heat. For the sake of simplicity, the influence of this
heating is not taken into account here, and dissipation acts solely as
an energy sink. By nonlinear transfer, energy-containing modes
generate a mean spectral flux of energy along its gradient in
wave number space while striving for energetic equipartition. This is
observed in an ideal system decoupled from external influence and
dissipation, see e.g. \cite{Lee1952}. Hence, the averaged interaction of
two wave numbers $\mathbf{k}$ and $\mathbf{q}$ tends to decrease
existing gradients. This defines a flux direction encoded by
\begin{align}
s_{kq}=\text{sgn}(\vert \mathbf{z}^\pm(\mathbf{q})\vert -\vert \mathbf{z}^\pm(\mathbf{k})\vert)
\end{align}
defining the increase of energy in $\mathbf{k}$ for $s_{kq}=1$ and
corresponding decrease in $\mathbf{q}$ or vice versa for
$s_{kq}=-1$. The actual information of the energy transfer direction
is encoded in the phase of the modes in the triad, and thus, remains
unknown if only the corresponding amplitude or energy is given, which
is the set of variables we choose based on our initial hypothesis.
For an analysis of the mean geometric characteristics of the transfer function in Eq. \eqref{els2}
with regard to the involved wave vectors,
we assume that all states $\mathbf{z}$ are equally likely.
By averaging over all possible orientations of the fields entering the transfer function the two contributions that have been identified above yield on average:
\begin{align}\label{geo1}
\vert \mathbf{k} \cdot \mathbf{z}^\mp(\mathbf{p}) \vert &\sim \frac{\Vert \mathbf{p}\Vert^2 \Vert \mathbf{k}\Vert^2-(\mathbf{k}\cdot \mathbf{p})^2}{\Vert \mathbf{p}\Vert \Vert \mathbf{k} \times \mathbf{p}\Vert}=:c_{kp}\,, \\
\vert \mathbf{z}^\pm (\mathbf{q}) \cdot \overline{\mathbf{z}^\pm (\mathbf{k})} \vert^2 &\sim \frac{1}{4}+(\frac{\Vert \mathbf{k}\Vert^2 \Vert \mathbf{q}\Vert^2(\mathbf{q}\cdot\mathbf{k})-(\mathbf{k}\cdot\mathbf{q})^3}{2 \Vert \mathbf{q}\Vert  \Vert \mathbf{k}\Vert  \Vert \mathbf{q}\times\mathbf{k}\Vert^2})^2=:e_{kq}^2\,.\label{geo2}
\end{align}
These geometric interaction factors for the two terms are closely
related to the triad interaction coefficients calculated in
\cite{Waleffe1992} using a helical decomposition.

The averages are attained by integration over all possible configurations of normalized vectors $\mathbf{z}$. Taking two vectors of length one (representing real and imaginary part of $\mathbf{z}$) and rotating them by all angles $\alpha$ by the rotation operator $\mathbf{R}_\mathbf{p}$ around their respective wavevector $\mathbf{p}$ while considering all attainable length ratios $r$ of their real and complex components, we derive, e.g. for the averaged absolute timescale
$\mathbf{k}\cdot\mathbf{z}(\mathbf{p})$:
\begin{widetext}
	\begin{align}\label{averot}
	\vert \mathbf{k}\cdot \mathbf{z}(\mathbf{p}) \vert \sim \frac{1}{4\pi^2} \int_0^{1}\int_0^{2\pi}\int_0^{2\pi} \sqrt{(\mathbf{k} \cdot r(\mathbf{R}_p(\alpha)(\frac{\mathbf{k} \times \mathbf{p}}{\Vert \mathbf{k} \times \mathbf{p} \Vert}))^2 +(\mathbf{k} \cdot \sqrt{1-r^2} (\mathbf{R}_p(\alpha')(\frac{\mathbf{k} \times \mathbf{p}}{\Vert \mathbf{k} \times \mathbf{p} \Vert}))^2}  \text{ d$\alpha$ d$\alpha'$ dr},
	\end{align}
\end{widetext}
where $\frac{\mathbf{k}\times \mathbf{p}}{\Vert \mathbf{k} \times \mathbf{p} \Vert}$ defines a normalized vector eligible to be rotated around its orthogonal vector $\mathbf{p}$ as $\mathbf{k}$ and $\mathbf{p}$ are not co-linear due to the triad constraint. The integral for the second interaction factor is derived similarly and both are straightforwardly evaluated by standard algebra. 

Combining the considerations of the triad geometry, the
gradient-diffusion approximation (GDA), and the proportionality of the
energy transfer function with the velocities of two active modes ($\mathbf{k}$ and $\mathbf{q}$) and one
passive mode ($\mathbf{p}$) defines the gradient diffusion transfer function  defined as the conservative change of energy of mode $\mathbf{k}$ with $\mathbf{q}$:

\begin{align}
	T_{GDA}^\pm(\mathbf{k},\mathbf{q}|\mathbf{p})= s_{kq} c_{kp}e_{kq} \vert \mathbf{z}^\mp(\mathbf{p})\vert\vert \mathbf{z}^\pm(\mathbf{q})\vert\vert \mathbf{z}^\pm(\mathbf{k})\vert,\label{transfer}
\end{align}
where the timescale of the transfer is given by $(c_{kp}\vert
\mathbf{z}^\mp(\mathbf{p})\vert)^{-1}$. Switching the roles of
$\mathbf{k}$ and $\mathbf{q}$ switches sign of the flux direction
while the geometric interactions coefficients are unchanged, implying
detailed conservation of energy. As a result, this energy-reduced
triad interaction is reformulated in port-Hamiltonian form:
\begin{align}
\partial_t \vert\mathbf{z}^\pm (\mathbf{k})\vert^2&=  T_{GDA}^\pm(\mathbf{k},\mathbf{q}|\mathbf{p})\\
\partial_t \vert\mathbf{z}^\pm (\mathbf{q})\vert^2&=  T_{GDA}^\pm(\mathbf{q},\mathbf{k}|\mathbf{p})\\
\partial_t \vert\mathbf{z}^\mp (\mathbf{p})\vert^2&=0 
\end{align}
with Hamiltonian function
\begin{align}
H(\mathbf{z})=\vert \mathbf{z}^\pm (\mathbf{k})\vert^2 +\vert \mathbf{z}^\pm (\mathbf{q})\vert^2 +\vert \mathbf{z}^\mp (\mathbf{p})\vert^2,
\end{align}
where $\mathbf{z}=(\vert \mathbf{z}^\pm (\mathbf{k})\vert^2 ,\vert \mathbf{z}^\pm (\mathbf{q})\vert^2 ,\vert \mathbf{z}^\mp (\mathbf{p})\vert^2) $ and
\begin{align}
J(\mathbf{z})=
\begin{bmatrix}
0 & T_{GDA}^\pm(\mathbf{k},\mathbf{q}|\mathbf{p}) & 0 \\
-T_{GDA}^\pm(\mathbf{k},\mathbf{q}|\mathbf{p}) & 0 & 0 \\
0 & 0 & 0
\end{bmatrix}.\label{EJ}
\end{align}
Dissipative effects are introduced similarly to \eqref{eq:diss} while the effect of a mean magnetic field requires additional modelling and physical insights as its influence is
indirect with regard to energy transfer and more complex.

The derivation has considered only the Els\"asser energies and, for
simplicity, has neglected the effects of finite magnetic helicity \cite{moffat:maghel}. However,
the influence of cross-helicity, i.e. the difference of the two Els\"asser
energies, is automatically and consistently accounted for by the
model. This is seen analytically by
\begin{align}
E^\pm&=\frac{1}{2}\int \frac{1}{2}(v^2+b^2) \pm \mathbf{v}\cdot \mathbf{b} \text{ dV}\\
&=\frac{1}{2}\int \frac{1}{2}(v^2+b^2) (1 \pm \sigma_r) \text{ dV}\\
&=\frac{1}{2}E^T(1\pm \sigma_r),
\end{align}
where $\sigma_r=\frac{H^C}{E^T}$ is the alignment, yielding
\begin{align}
z^\pm=\sqrt{\frac{1}{2}E^T(1\pm\sigma_r)}.
\end{align}

Hence, with stronger alignment $\sigma_r$, one of the Els\"asser energies
dominates and diminishes the nonlinear cascade favouring further
alignment due to the two-to-one correspondence of the Els\"asser velocities in the transfer function, see \eqref{transfer}.

In theoretical considerations of MHD turbulence, different physical
processes are underlying the nonlinear interaction of the turbulent
fluctuations.  These are characterized by specific
timescales, such as advective shearing on the nonlinear turnover time, $\tau_\text{NL}\sim \ell/v_\ell$, with $v_\ell$ a characteristic velocity on
scale $\ell$. Depending on the considered regime of turbulence, e.g.,
strong anisotropy due to a mean magnetic field or weak wave turbulence
involving Alfv\'en wave collision as prevailing interaction mechanism \cite{Kraichnan65},
different nonlinear transfer rates of ideal invariants are
predicted, e.g. involving the Alf\'en time, $\tau_\text{A}=\ell_\parallel/b_0$ where $\ell_\parallel$ denotes the characteristic extent of the
wave pulses along the magnetic field and $b_0$ is identical to the Alfv\'en speed in properly chosen Alfv\'enic units. Assuming a constant flux of energy, those predictions
result in characteristic self-similar inertial-range scaling laws for,
e.g. the energy spectra. As the modelling of the influence of the mean
magnetic field requires such additional complications, we adapt the
change of timescale to our energy transfer function accordingly. We
differentiate between the parallel and orthogonal components of the
triad wavevectors with respect to the mean field direction, indicated
by the subscripts $\perp$ and $\parallel$, respectively. The
Els\"asser variables characterize Alfv\'en wave pulses traveling
along the magnetic field lines with Alfv\'en speed  that
mutually collide and interact nonlinearly on the timescale defined by their extent along the
magnetic-field direction and their speed, where the extent is
estimated  as $\ell_\parallel \sim k^{-1}_\parallel$. As the mediator defines the
timescale in our model, it also serves for the purpose of
phenomenological adaptation defined by
\begin{align}
\tilde{c}_{kp}=c_{kp}\frac{\vert \mathbf{p}_\perp \vert }{\vert \mathbf{p}_\parallel \vert}\frac{\vert \mathbf{z}^\mp(\mathbf{p})\vert}{\vert \mathbf{b_0} \vert}\label{tauadapt}\,.
\end{align}
We choose to customize the geometrical interaction factor which is
part of the timescale definition. The physical dimension remains
unchanged while the interaction strength is adapted according to the
considered mechanism. We discern two scenarios: in the weak
turbulence regime, mutual Alfv\'en wave scattering is considered as
dominant energy transfer mechanism, which either occurs in a system
without a mean magnetic field where Alfv\'en waves travel along the
root-mean-square field of the largest and slowliest evolving magnetic
fluctuations, or in a system, where a strong, external mean field acts
on the dynamics. In the first case, parallel and orthogonal extent of
the smallest excited wave numbers are assumed to be equal, changing the
timescale by the quotient of Els\"asser velocity and Alfv\'en
speed. For the second, a more prounounced dynamical anisotropy is present: The orthogonal
dynamics are unaffected by the mean field while the parallel dynamics exhibit an increase of the flux timescale. To introduce the directional
dependence of the interaction
rate accordingly, the modification of the geometrical interaction factor is given by
\begin{align}
\tilde{c}_{kp}=c_{kp}\min (1,\frac{\Vert \mathbf{p}_\perp \Vert }{\Vert \mathbf{p}_\parallel \Vert}\frac{\Vert \mathbf{z}^\mp(\mathbf{p})\Vert}{\Vert \mathbf{b_0} \Vert} )
\end{align}
which includes the effect of the mean field while respecting that the
cascade time cannot be shorter than the wave period, i.e., the
modification factor cannot surpass one.
\section{Numerical Experiments}
The gradient diffusion network is initialized by the definition of
wave number nodes, their connections, and initial Els\"asser
energies. According to the discrete radii $r_n=r_0^n$ and
cylindrical/spherical angles $\alpha_n=\frac{2\pi n}{N}$ of
a spherical or cylindrical Fourier discretization,
wave number space is sparsely filled. The geometric interaction factor
favors local energy fluxes, i.e., between wave numbers of similar size,
which is observed in direct numerical simulations and theoretically
expected by analytical considerations \cite{Kraichnan1971}. As a
consequence, only modes up to a limiting \lq local neighbourhood
distance\rq\ are connected to further reduce model complexity as no
significant changes of the results are observed when adding non-local, i.e. distant,
connections. We define the distance between nodes by
\begin{align}
d_{loc}=\vert \log_s(\vert \mathbf{k}\vert)-\log_s(\vert \mathbf{q}\vert)\vert.
\end{align}
The process of connecting two nodes requires a mediator node, which is
derived by minimizing the error of the triad constraint,
$\mathbf{k}+\mathbf{p}+\mathbf{q}=0$. Due to the intrinsic detailed
conservation, the triad relation does not need to be fulfilled exactly
although its influence on the involved wave numbers and timescales
needs to be accounted for.  For the numerical experiments conducted
in the following, node networks in spherical or cylindrical geometry
are generated with geometrically increasing factor $r_0=1.4$,
connections are selected with $d_{loc}\leq4$ with $s=2$, and the
number of discrete angles is set to $N=6$. Variations of the number of
angles and factor $r_0$ do not alter the general behavior of the model
and are, therefore, not discussed specifically in the following.

A considerable range of wave numbers is covered as compared to direct
numerical simulations by a number of radial shells set to $N_r=40$,
yielding a maximum wave number of $k_\text{max}\sim 10^6$ that allows
to set the values of viscosity and resistivity to
$\mu=\nu=10^{-7}$. Due to the energy based modelling approach, all
nodes are required to contain energy to actively participate in the
system's dynamics. Hence, the network is initialized with Els\"asser
energies according to a $k^{-5/2}$-spectrum which is significantly
steeper than any observed turbulent spectrum to easily discern any effect
of the initial conditions. A statistically stationary state is
obtained by replenishing dissipated energy via a forcing method that
enforces constancy of the energies carried by the network nodes in the
first wave number shell of smallest radius.

In the following, we show the general functionality of the model and
its timescale parameters while exploring some of its limits that hint
to further extensions and developments.

\subsection{Basic Gradient Diffusion Network}
The network model with an interaction time not modified by the presence of magnetic fields
corresponds to strong turbulence, i.e. the timescale of unit nonlinear interaction, $\tau_{NL}$, is comparable of
the timescale governing the cross-scale energy flux. Due to conservation of the respective Elsaesser
energies in the case of  statistical stationarity of the system,
the temporal averages of the ideal nonlinear flow rate and the energy dissipation rate, $\varepsilon^\pm$, are identical.

\begin{figure}[]
	\
	\begin{center}
		\includegraphics[width=\linewidth]{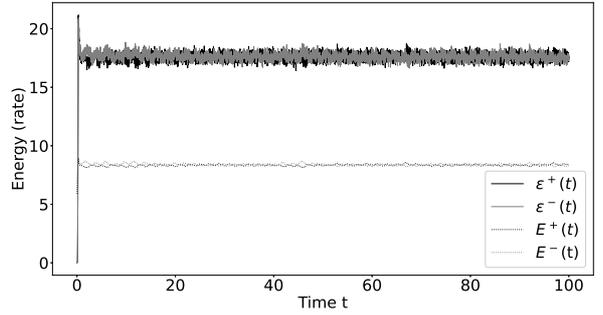}

	\end{center}
	\caption{Els\"asser energies $E^\pm$ (bottom) and dissipation rates $\varepsilon^\pm$ (top) with respect to simulation time for spherical shell coordinates of the gradient diffusion network.}\label{K41Energy}
\end{figure}

The visible quasi-stationary temporal oscillation of the
dissipation/flux rates and Elsaesser energies shown in Fig. \ref{K41Energy} are remnants of the initial conditions. Uniformly-distributed random
factors from the interval $[-1,1]$ are multiplied with each mode outside the
forcing range in the initialization step, which causes significant fluctuations of
energy fluxes between modes embodied in energetic fluctuations on all
scales. Thus, the total energy dynamics, dominated by the
small wave number modes, also fluctuate visibly.
Dropping the random
perturbations in the initial conditions eliminates this effect.
The quasi-stationarity these perturbations demonstrates the model's robustness and stability.

\begin{figure}[]		
	\begin{center}
		\includegraphics[width=\linewidth]{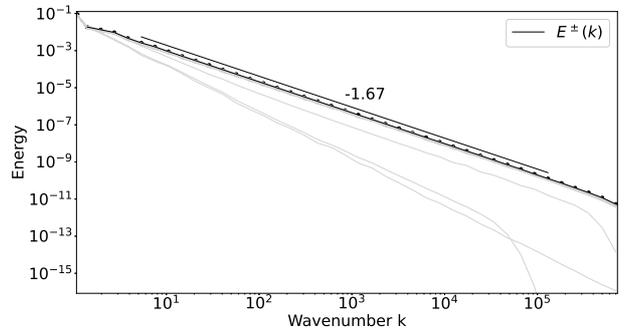}

	\end{center}
	\caption{Averaged energy spectrum $E^\pm(k)$ (black) of the gradient diffusion network in spherical shell coordinates with single time-instances in gray showing the evolution from initial conditions towards a statistically steady state.\label{K41spectrum}
} 
\end{figure}

After reaching a quasi-steady state the shell-integrated energy spectrum
averaged over the quasi-stationary period (see Fig.~\ref{K41spectrum})
displays perfect Kolmogorov scaling, $\sim k ^{-5/3}$ with little
variations for specific instances in time.  The curves in light gray
show how starting from the initial dissipation dominated spectrum, the
direct energy cascade subsequently fills the energy reservoirs at
higher wavenumbers in a self-similar fashion.

\begin{figure}[]		
	\begin{center}
		\includegraphics[width=\linewidth]{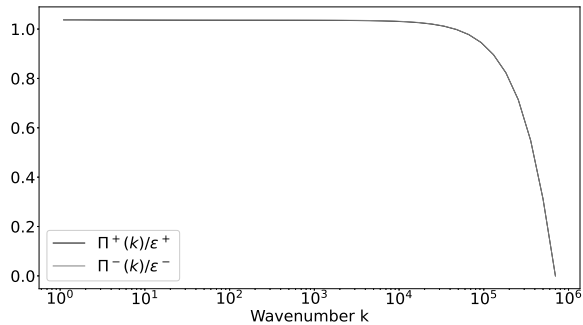}

	\end{center}
	\caption{Averaged cross-scale energy flux $\Pi^\pm(k)$ of the gradient diffusion network in spherical shell coordinates normalized by the respective
	time-averaged  dissipation rate~$\varepsilon^\pm$. \label{K41flux}}
\end{figure}

The time averaged cross-scale fluxes of the Elsaesser energies 
$\Pi^\pm(k)=\sum_{\vert\mathbf{r}\vert >k,\vert\mathbf{q}\vert <k}
T^\pm_{GDA}(\mathbf{r},\mathbf{q}|\mathbf{p})$ across wave number $k$ 
with respect to time and normalized by the respective dissipation rates are depicted in Fig. \ref{K41flux}. In the scaling range,
they are close to the respective dissipation rates
which results in quasi-stationary spectra of the Els\"asser energies
$E^\pm$. The slight deviation from unity is a consequence of our simple definition of $\Pi^\pm(k)$ which does not yield a energy flow strictly along
the radial direction in network space, but also includes oblique contributions.
Due to Elsaesser equipartition, cross-helicity, $H^C=(E^+-E^-)/4$ is absent in this system.	


\subsection{Mean magnetic field influence}
\begin{figure}[]
	\
	\begin{center}
		\includegraphics[width=\linewidth]{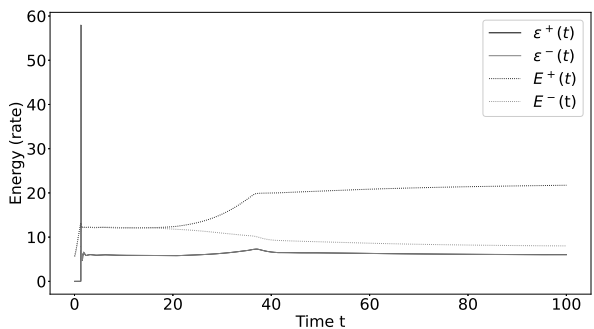}

	\end{center}
	\caption{Els\"asser energies $E^\pm$ (bottom) and dissipation rates $\varepsilon^\pm$ (top) with respect to their temporal evolution for spherical shell coordinates according to an energy flux timescale expressing mutual Alfv\'en-wave scattering (see Eq.~\eqref{tauadapt}).}\label{IKEnergy}
\end{figure}

\begin{figure}[]		
	\begin{center}
		\includegraphics[width=\linewidth]{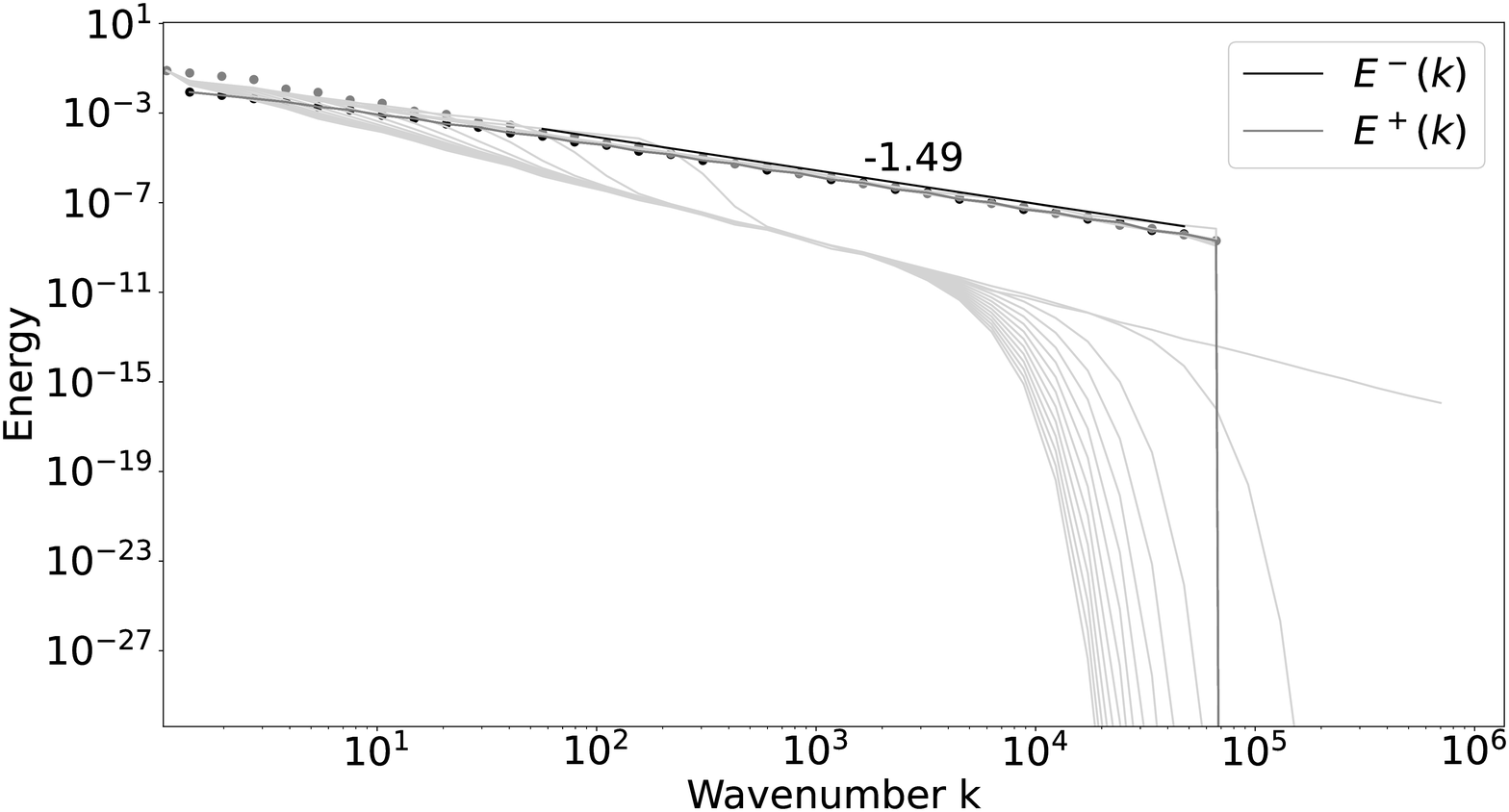}
	\end{center}
	\caption{Averaged energy spectrum $E^\pm(k)$ (black) of the gradient diffusion network in spherical shell coordinates according to
	an energy flux timescale expressing mutual Alfv\'en-wave scattering (see Eq.~\eqref{tauadapt}) with single instances in gray showing the evolution from initial conditions towards a statistically steady state.}\label{IKspectrum}	
\end{figure}

\begin{figure}[]		
	\begin{center}
		\includegraphics[width=\linewidth]{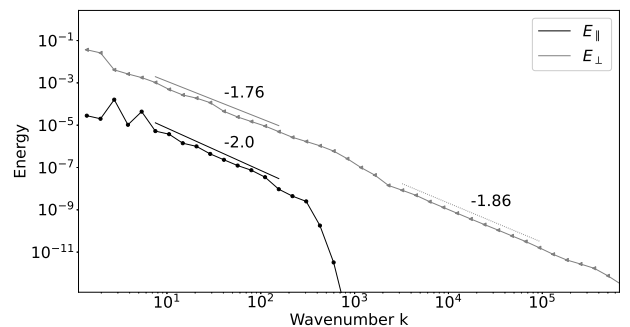}
	\end{center}
	\caption{Averaged anisotropic energy spectrum $E_{\parallel, \perp}(k)$ of the gradient diffusion network in cylindrical shell coordinates according to a flux timescale incorporating the anisotropic influence of an externally imposed mean magnetic field.}\label{GSspectrum}	
\end{figure}

\begin{figure}[]		
	\begin{center}
		\includegraphics[width=\linewidth]{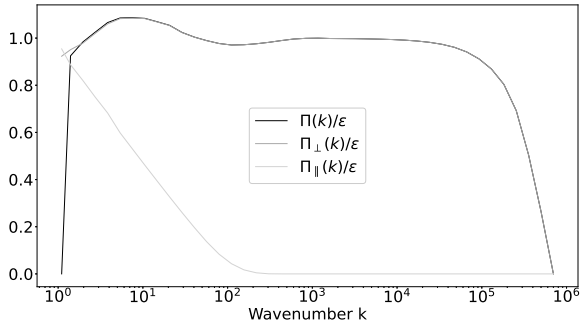}
	\end{center}
	\caption{Averaged cross-scale energy fluxes $\Pi_{\parallel, \perp}(k)$ of the gradient diffusion network in cylindrical shell coordinates according to a flux timescale incorporating the anisotropic influence of an externally imposed mean magnetic field normalized by the average dissipation rate $\varepsilon$.}\label{GSfluxes}	
\end{figure}

The inclusion of the anisotropic influence induced by a homogeneous mean magnetic on the nonlinear dynamics is carried out by the physically
motivated modification of the transfer timescale (see Eq.~\eqref{tauadapt}). This yields quasi-stationary power-law spectra with
a different scaling behaviour compared to first numerical experiment described above. Initializing large-scale spectral energetic isotropy results in 
spectral scaling very close to an Iroshnikov-Kraichnan-like spectrum \cite{Kraichnan65,Irsoshnikov63}, $~k^{-3/2}$, shown in  Fig. \ref{IKspectrum}.
The system exhibits constant averaged cross-scale flux, with an earlier dissipative fall-off
as compared to the previous strong turbulence case, due to the weaker interaction and,
consequently, a reduced nonlinear flux.
Interestingly, Fig. \ref{IKEnergy} displays some spontaneous self-organization of the system with an emerging finite difference
of the Elsaesser spectra, i.e. finite cross helicity due to alignment of $\mathbf{v}$ and $\mathbf{b}$.
In the spectrally anisotropic
large-scale setup, on the other hand, more complex dynamics evolve. To
enhance the spectral resolution in the direction of the mean magnetic
field, a cylindrical coordinate system is considered, with an  increased
number of field-parallel nodes in the direction of the external
mean field is chosen. Theoretically and neglecting the difference between globally and locally orthogonal
directions with respect to $\mathbf{b}$, a Kolmogorov
spectrum is expected along the field-orthogonal direction and a weak
turbulence spectrum, $\sim k_\parallel^{-2}$, along the field-parallel direction.
Although the spectra display slight variations, see Fig.~\ref{GSspectrum},
the general behavior is captured and deviations can
be explained by the model geometry. Due to the very sparse grid, the
parallel and orthogonal interactions partially decouple in the low
wave number region. However, the cylindrical shells that represent the
spectral energy for a specific $k_\bot$ extend along the
field-parallel direction over all wave numbers. As the parallel energy
spectrum falls off at much lower parallel wave numbers than the
orthogonal spectrum, this spectral anisotropy creates a non-homogenous
set of active energy-containing nodes in the cylindrical shells
determining the orthogonal energy spectrum.  Consequently, the
resulting spectral transfer function of our reduced model exhibits spectral variations,
responsible for the irregularity in the orthogonal spectrum at the end
of the approximate parallel scaling range. The pronounced anisotropy of the spectral energy flux
is clearly shown in Fig. \ref{GSfluxes}. 
\begin{figure}[]		
	\begin{center}
		\includegraphics[width=\linewidth]{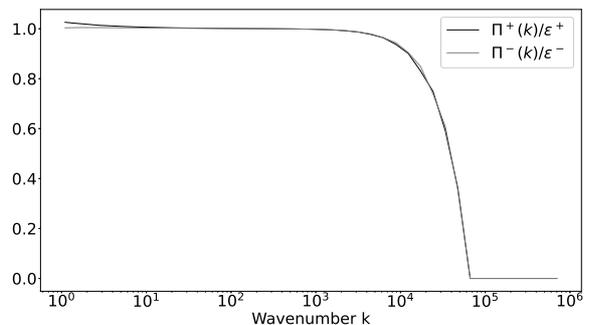}

	\end{center}
	\caption{Averaged cross-scale energy flux $\Pi^\pm(k)$ of the gradient diffusion network in spherical shell coordinates according to an the energy flux timescale expressing mutual Alfv\'en-wave scattering (see Eq.~\eqref{tauadapt}) normalized by the average dissipation rate $\varepsilon^\pm$.}\label{IKflux}	
\end{figure}

\subsection{Decaying turbulence}

\begin{figure}[]		
	\begin{center}
		\includegraphics[width=\linewidth]{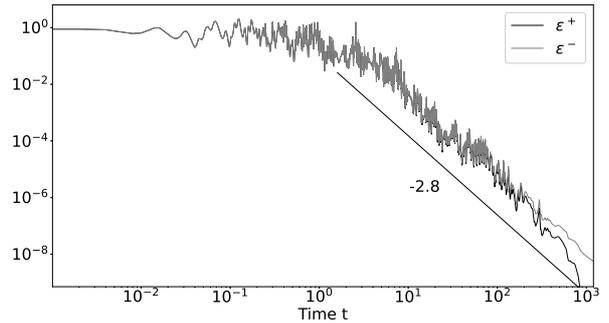}
	\end{center}
	\caption{Decay of the dissipation rates $\varepsilon^\pm=\partial_t E^\pm(t)$ of the gradient diffusion network approximately indicating a $E^\pm(t)\sim t^{-1.8}$ power-law.}\label{Decaylog}	
\end{figure}

An unforced gradient diffusion network establishes a Kolmogorov
spectrum similar to the forced case while its total energy diminishes
according to a power-law $E(t)\sim t^{-\beta}$
\cite{Kolmogorov1941,Saffman1967,biskamp:book,BiskampMuller1999}. In
contrast to hydrodynamic Navier-Stokes decay, the MHD case is reported to be
significantly shallower due to the different structuring of the flow imposed by the additional magnetic field, i.e.,
$\beta \in [\frac12,1]$ \cite{biskamp:book,BiskampMuller1999,Biskamp2000} (MHD)
rather than $\beta \in
[1,2]$\cite{Hinze,Mohamed1990,Smith1993} (Navier-Stokes). Figure \ref{Decaylog}
reflects the behavior expected by hydrodynamic turbulence
($t^{-1.8}$), although the interaction of Els\"asser variables is
considered. Therefore, it is apparent that the model lacks dynamics
likely related to magnetic helicity and its property to generate and preserve large-scale flow structures.
The two Els\"asser fields solely advect each other capturing nonlinear alignment effects and linear Alfv\'enic dynamics;
the process of  magnetic structure formation is not present in the current version of our reduced network model.
This is an evident explanation of 
the different decay characteristics of the model as compared to direct numerical
simulations (cf., e.g., \cite{BiskampMuller1999}). Additionally, the universality of energy decay is in
question as the decay exponent might be dependent on the
configuration of the large-scale characteristics of the flow, which are hardly captured by the reduced model. The
model, however, is not expected to be able to capture such complex
dynamics yet, it rather offers a starting point to incorporate more
sophisticated dynamics in its modular structure.

\section{Conclusion}

The present gradient-diffusion network model of magnetohydrodynamic turbulence
combines phenomenological considerations and geometrical analysis of
the nonlinear energy transfer function into a reduced-order representation of
conservative spectral transport of
energy and cross-helicity, which is formulated in port-Hamiltonian form.

Motivated by the fundamental Els\"asser symmetry of the MHD equations, the dynamic
network representation is based on an abstraction of the  nonlinear triad interactions of MHD as a sum of conservative pair-exchanges
of like-signed Els\"asser energies.
The third member of any interacting triad while not actively participating in the exchange of Els\"asser energies
takes the role of a mediator by setting the timescale of the transfer. An analysis of the mathematical
structure of the exact Fourier transfer function allows to identify, next to the influence of the mediator,
the important role played by the wave vector geometry of each triad for the amount of transferred energy.
With these insights, the energy dynamics of incompressible turbulent flow can be modeled via a network
made up of nodes corresponding to Fourier modes carrying two Els\"asser energies.
This reduced representation of turbulence has the general characteristics of a port-Hamiltonian model.
Its modular structure allows straightforward and flexible changes to the size and shaping of the network which
is a significant advantage compared to classical shell-models. Furthermore, the network ansatz has no limitations
with regard to spatial dimensions. It is therefore not difficult to model anisotropic configurations that
occur due to turbulent magnetic fields. The principal control parameters of the model are the dominant timescales of
nonlinear interaction.
Due to the sparse discretization of Fourier space that is possible in the network architecture,
the model can attain a spectral bandwidth and associated Reynold numbers which are comparable to classic shell-models, however,
without sharing their structural restrictions.

Numerical experiments testing the basic properties such as consistency, and stability, of the model system
show stationary self-similar inertial-range solutions of MHD energy
spectra  and spectral fluxes in agreement with theoretical expectations and results from direct numerical simulations.
By simple extension of the set of underlying timescales, the model also allows to correctly
reproduce anisotropic behaviour expected in systems permeated by a constant mean magnetic field.
Simple tests with regard to turbulence decay also gave results consistent with present knowledge.
 
Although these encouraging results  demonstrate
the basic functionality and consistency of the model,
some fundamental dynamic properties of incompressible turbulent MHD  still need to be included
in the reduced representation. An important next step towards a more comprehensive
physical representation is the inclusion of the third ideal invariant of incompressible
MHD, magnetic helicity, $H^M=1/(2V)\int \mathrm{d}V \mathbf{a}\cdot\mathbf{b}$, with the
magnetic vector potential $\mathbf{a}$, $\mathbf{b}=\nabla\times\mathbf{a}$.
This quantity is responsible for the formation of large-scale magnetic structures from small-scale
MHD turbulence. As such large-scale structures have a profound effect on turbulent dynamics and energetic
modelling of magnetic helicity will be the next step of development and is currently under way.

\begin{acknowledgements}
The authors thank V. Mehrmann for fruitful discussions.
This work was supported by the German Research Foundation (DFG) within the Research Training Group GRK2433.
\end{acknowledgements}

\bibliographystyle{plain}
\bibliography{references.bib} 

\end{document}